\documentclass[12pt, onecolumn, draftcls,]{IEEEtran}

\usepackage[english]{babel}
\usepackage[T1]{fontenc}
\usepackage{wrapfig}
\usepackage{inputenc}
\usepackage{cite}
\usepackage{color}

\usepackage{enumitem}

\usepackage[cmex10]{amsmath}
\usepackage{amssymb}
\usepackage{amsthm}

\usepackage{array}
\usepackage{multirow}
\usepackage{colortbl}
\usepackage{soul}

\usepackage{graphicx}
\usepackage{epstopdf}
\usepackage{multirow}

\usepackage{booktabs}
\usepackage{url}
\usepackage{hyperref}
\usepackage[nolist]{acronym}
\usepackage{bibentry}

\definecolor{gold}{rgb}{0.85,.66,0}

\newcommand{\abs}[1]{\left\vert {#1} \right\vert}

\newcommand{\real}[1]{\text{Re}\left\lbrace {#1} \right\rbrace}
\newcommand{\imag}[1]{\text{Im}\left\lbrace {#1} \right\rbrace}

\newcommand{\mt}[1]{ {\textsc{{#1}}} }

\newcommand{\rbrac}[1]{\left( {#1} \right)}
\newcommand{\sbrac}[1]{\left[ {#1} \right]}
\newcommand{\cbrac}[1]{\left\lbrace {#1} \right\rbrace}
\newcommand{\ceil }[1]{\left\lceil  {{#1}} \right\rceil}
\newcommand{\floor}[1]{\left\lfloor {{#1}} \right\rfloor}

\begin{document}

\title{Closed-Form Bit Error Probabilities for FBMC Systems}
\author{Ricardo Tadashi Kobayashi and Taufik Abrao\\
\normalsize Department of  Electrical Engineering (DEEL)
State University of Londrina  (UEL). \\
Po.Box 10.011, CEP:86057-970, Londrina, PR, Brazil. \\
 Email: ricardokobayashi@uel.br; taufik@uel.br

\thanks{R. T. Kobayashi and T. Abrao are with the Department of  Electrical Engineering (DEEL)
State University of Londrina  (UEL). Po.Box 10.011, CEP:86057-970, Londrina, PR, Brazil.  Email: ricardokobayashi@uel.br; taufik@uel.br}
}
\maketitle

\begin{abstract}
This paper  analyses the data reconstruction effects  emerged from the deployment of non-perfect prototype filters in \ac{FBMC} systems operating over \ac{AWGN} and {frequency-flat Rayleigh channels considering frequency-flatness for each subcarrier}. This goal is attained by studying the \ac{BER} effects of the prototype filters, increasing the scenario complexity progressively. Despite the  complexity, both exact and approximate \ac{BEP} expressions {portray} the \ac{BER} degradation analytically for any \ac{FBMC} prototype filter. Numerical results  demonstrate that the proposed {\ac{BEP}} expressions match perfectly with the simulated \ac{BER} performance for \ac{FBMC} systems, regardless of the prototype filter choice.
\end{abstract} 
\acresetall
\smallskip
\noindent 

\section{Introduction}\label{sec:fbmc_intro}
With the ever-growing popularity of wireless communications, services {have} become increasingly complex and demanding in order to meet the stringent user requirements.  Therefore, \ac{5G} compliant technologies is a hot research topic at this point \cite{Shafi2017,Wang2018}. Among them, we can name massive \ac{MIMO} systems, mmWaves, and mMTC as the most promising technologies for \ac{5G} \cite{Boccardi2014}. Massive \ac{MIMO} systems deploy a large number of antennas to enhance spectral/energy efficiency \cite{Nadeem2018}, while mmWaves devices operate in frequencies over 30 GHz \cite{Rappaport2017}, where spectrum is abundant. Despite not being expected to be fully supported, basic concepts of Cognitive Radio can be useful for \ac{5G} \cite{Sasipriya2016} and other wireless systems to come, as spectrum usage can be considerably enhanced. Another aspect of utmost importance is the radio access, as spectral efficiency is primordial for dense networks.

\ac{OFDM} is a classical waveform, being deployed in contemporary standards such as 802.11 for local networks, 802.16 for Wimax and LTE-A for 4G systems \cite{Hwang2009}. Despite its popularity, such waveform  presents some limitations. First, \ac{OFDM} is known to present high \ac{OoB} emission due to its rectangular envelope. Moreover, \ac{OFDM} channel equalization requires the usage of reasonably long \ac{CP}, reducing the spectral efficiency. Another drawback is the reliance of \ac{OFDM} on its orthogonality, which prevents its efficient operation for random access channel. In order to address such issues, alternative multicarrier schemes have been intensively studied in the last years \cite{Wunder2014}.

Among the waveforms alternatives, \ac{FBMC} is considered one of the most promising radio access for wireless systems to come {\cite{Farhang2011,Perez2016,Cai2018,Farhang2014b}}. One of the reasons for such prospect is {the} robustness of \ac{FBMC} systems against \ac{ISI} without the usage of long \acp{CP} \cite{Zhang2017}. According to \cite{Farhang2014}, \ac{FBMC} systems present a synergistic operation with massive \ac{MIMO}. Furthermore, \ac{FBMC} systems are also known to provide very low \ac{OoB} emission, favoring spectrum efficiency. Hence, such aspects made \ac{FBMC} one of the waveforms selected for the METIS project \cite{Metis} and the main choice for PHYDIAS project \cite{Phydias}.

Although \ac{FBMC} systems have been extensively studied in the last few years, they are still much less explored than \ac{OFDM}, and there are still many open issues to be addressed. Despite their similarities, \ac{OFDM} methods and algorithms are not fully compatible with \ac{FBMC} systems  as they rely on orthogonality in the real field. Hence, channel estimation is proceeded differently due to the imaginary interference emerged in \ac{FBMC} systems \cite{Cui2016}, and \ac{PAPR} methods such as the Tone Reservation also needs to be modified \cite{Rahmatallah2013,Bulusu2015}. 

The prototype filter choice is yet another subject to be addressed, as they control the overall spectrum and reconstruction features of \ac{FBMC} systems. Despite the  Mirabbasi-Martin filter  \cite{Martin1998,Mirabbasi2003} being the most widespread prototype filter, there are plenty other options \cite{Sahin2014,Prakash2013,Bellanger2001,Siohan2000,Vahlin1996,Moore2004,Nadal2018,Aminjavaheri2017,Kobayashi2019} that offer different \ac{OoB} emission, time/frequency pulse dispersion, and \ac{SIR} trade-offs.

As the deployment of different prototype filter yields different \ac{SIR} levels, symbol reconstruction performance, \textit{i.e.}, \ac{BER} performance is expected to be different. Despite the effects of the prototype filter on the overall performance of  \ac{FBMC} transmissions being extensively studied, the provision expressions for \ac{BEP} have been ignored so far. Such subject remained ignored as prototype filters can be designed to achieve large \ac{SIR} levels, meaning marginal effects on the \ac{BER} performance. Moreover, one can also {deploy} \ac{SIR} or distortion measurements to assess the effectiveness of the prototype filter on the overall transmission of an FBMC system qualitatively. However, beyond the aforementioned facts, the provision of closed-form \acp{BEP} has not been made yet due to their non-trivial solution. From this perspective, the {\it main contribution of this paper lies in the provision of analytic expressions of \ac{BEP} for \ac{FBMC} systems, regardless of the prototype filter option}. Hence, as an effective contribution and novelty, through the results of this work, {\it we quantify the prototype filter effect on the \ac{BER} performance of an \ac{FBMC} system}, which was known only qualitatively until now.

The rest of the paper is organized as follows. Section \ref{sec:fbmc_mux} presents the basics of \ac{FBMC} systems. Section \ref{sec:Pe-SC} discusses {the foundations} for evaluating \acs{BEP} in {both} \ac{AWGN} and {frequency-flat} Rayleigh channels. In Section \ref{sec:analytic}, analytic formulas for evaluating the \ac{BEP} of \ac{FBMC} systems are proposed. Section \ref{sec:results} presents corroborative numerical results {that compare the numerical \ac{BER} with the derived \ac{BEP} expressions. Finally,} Section \ref{sec:concl} closes the paper by offering conclusions and final remarks.
\vspace*{3mm}
\noindent\textbf{Notation:}  $\real{\cdot}$ and $\imag{\cdot}$ are the real part and imaginary part operators, while $j=\sqrt{-1}$. $\ceil{\cdot}$ is the ceil operator. {$\left< a[k] \left\vert b[k] \right.\right>=\sum_{k}  a[k] b^*[k]$ is the inner product between $a[k]$ and $b[k]$, where $(\cdot)^*$ is the complex conjugation operator.} {$u(\cdot)$ and $\delta{(\cdot)}$} are, respectively, the step function and the Dirac delta function.

\section{\ac{FBMC} Multiplexing}\label{sec:fbmc_mux}
The \ac{FBMC} multiplexing distributes phase-shifted \ac{PAM} symbols along $M$ subcarriers with a specific pulse shaping. The discrete representation of an \ac{FBMC} signal consists of the superposition of \ac{PAM} symbols shaped by their respective pulses, i.e.,
\begin{equation}
	s[k]  =  \displaystyle\sum_{n=-\infty}^{\infty} 
			 \displaystyle\sum_{m=0}^{M-1}
			   a_{m,n}  p_{m,n}[k],
	\label{eq:fbmc_siso_disc}
\end{equation}
where $a_{m,n}$ is the $n$th PAM symbol of the $m$th subcarrier and $p_{m,n}[k]$ is the associate pulse shape of length $L_p$\footnote{Typical values for $L_p$ are $KM-1$, $KM$ and  $KM+1$, where $K$ is the overlapping factor of the prototype filter.}, which is given by
\begin{equation}
	p_{m,n}[k] = p\sbrac{k-n\dfrac{M}{2}}
			  e^{ j\rbrac{\frac{2\pi}{M} m \underline{k} + \phi_{m,n}}},
\end{equation}
where $p[k]$ is the prototype filter, $\underline{k}=k-(L_p-1)/2$ and $\phi_{m,n}$ is the phase-shift {added} to $a_{m,n}$. Notice that $\phi_{m,n}$ is designed to make the adjacent symbols phase-shifted by $\pi/2$ and {typically set to}
\begin{equation}
	\phi_{m,n} = \dfrac{\pi}{2}(m+n),
	\label{eq:fbmc_phase1}
\end{equation}
which will be deployed throughout the remainder of this work.

\subsection{Symbol Reconstruction}
In order to recover the \ac{PAM} symbols at the receiver side, the received \ac{FBMC} signal must be correlated to the pulse of the desired symbol. Hence, {for a noiseless scenario,} the symbol $a_{m_0,n_0}$ can be retrieved by taking the real part of the $s[k]$ projection onto $p_{m_0,n_0}[k]$: 
\begin{equation}
	\tilde{a}_{m_0,n_0} =
	\real{{\left< s[k] \left\vert p_{m_0,n_0}[k] \right.\right>} }.
	\label{eq:fbmc_est1}
\end{equation}
Notice that $\cbrac{p_{m,n}[k]}$ is not strictly orthogonal, leading to symbol self-interference. {By expanding eq. \eqref{eq:fbmc_est1}}, one can evaluate the whole  interference experienced by the $n_0$th symbol of the $m_0$th subcarrier:
\begin{equation}
	\hspace*{-1mm}
	\tilde{a}_{m_0,n_0} = 
	a_{m_0,n_0} +
	\displaystyle\sum_{\substack{ n\neq n_0 \\m\neq m_0}}
	\hspace*{-1mm}
	a_{m,n} 
	\real{\left< p_{m,n}[k] \left\vert p_{m_0,n_0}[k] \right.\right>}. 
	\label{eq:fbmc_est2}
\end{equation}
Therefore, an appropriate prototype filter must be deployed in order to enable  near-perfect reconstruction at the receiver, since it introduces the self-interference highlighted in eq. \eqref{eq:fbmc_est2}. {Since the prototype filter choice can affect considerably symbol reconstruction, the characterization of this effect is very important and is investigated in this work through the provision of \ac{BEP} expressions that describe such {phenomenon}.}

{Indeed, prototype filters can be designed to achieve perfect reconstruction, meaning no distortion during symbol reconstruction described in eq. \eqref{eq:fbmc_est2}. In this sense, a prototype filter is said to offer perfect reconstruction if it satisfies eq. (21) of \cite{Nguyen1996}. Nevertheless, a small amount of distortion is almost always allowed to improve the spectral performance of prototype filters deployed in \ac{FBMC} systems \cite{Mirabbasi2003}.}

\subsection{Interference Elements}
As observed in eq. \eqref{eq:fbmc_est2}, the prototype filter introduces an interference on the symbol reconstruction procedure. Since this phenomenon degrades the \ac{BER} performance, let us define
\begin{equation}
	\arraycolsep = 1.4pt
	\hspace*{-2.5mm}
	\begin{array}{rcl}
		\epsilon_{m,n} & = & \real{\left< p_{m,n}[k] \left\vert p_{0,0}[k] \right.\right>}, \quad n,m \neq 0
\\[3pt]
					   & = & \cos\rbrac{ \phi_{m,n} } 
					         \hspace*{-1mm}
		                   	 \displaystyle\sum_{k=-\infty}^{\infty}
		                   	 \hspace*{-1mm}
							 p\sbrac{k-n\frac{M}{2}} p[k]
					         \cos\rbrac{ \dfrac{2\pi}{M} m  \underline{k}  }	\end{array}
	\hspace*{-3mm}
	\label{eq:fbmc_dist}
\end{equation}
as the interference introduced by the symbol $a_{m,n}$ into the symbol $a_{0,0}$.

From \eqref{eq:fbmc_dist}, one can enumerate the following properties for the interference elements:
\begin{itemize}
	\item[i.]   $\epsilon_{0,0}$ represents the pulse energy;
	\item[ii.]  $\epsilon_{m,n}$ is an even sequence concerning the index $n$, \textit{i.e.}, $\epsilon_{m,n}=\epsilon_{m,-n}$;
	\item[iii.] $\epsilon_{m,n}$ is even circular symmetric concerning $m$, \textit{i.e.}, $\epsilon_{m,n}=-\epsilon_{M-m,n}${, for $1\leq m \leq M/2$};
	\item[iv.]  $\epsilon_{m,n}=0$ for $\abs{n}>\ceil{\frac{L_p}{M/2}}-1$, since the length of the  prototype filter is finite;
	\item[v.]   $\epsilon_{m,n}=0$ case $m+n$ is odd, given the cosine term.
\end{itemize}
These five properties can be deployed to define 
\begin{equation}
	\mathcal{E} = 
	\cbrac
	{
	{\epsilon_{m,n}\,} \quad
	\left\vert
	\begin{array}{l}
		0\leq m \leq M-1 \\
		\abs{n} \leq \ceil{\frac{L_p-1}{M/2}}-1 \\
		m+n \text{ even} \\
		m+n\neq 0
	\end{array}
	\right.
	}
	\label{eq:interf_set}
\end{equation}
as the set containing all the non zero interference elements $\epsilon_{m,n}$.

It is noteworthy mentioning that the interference set $\mathcal{E}$ contains
\begin{equation}
\abs{\mathcal{E}} = 
 M\rbrac{ \ceil{\frac{L_p-1}{M/2}}-\dfrac{1}{2} } -1
\end{equation}
elements and is a keypoint  to derive the {\acp{BEP}} for \ac{FBMC} systems deploying arbitrary prototype filters.

\section{$N_p$-PAM {\ac{BEP}}: Single Carrier Case}\label{sec:Pe-SC}
In order to provide a comprehensive presentation of the subject,  let us present first the single carrier case for calculating the {\ac{BEP} of the transmission of \ac{PAM} symbols}. Initially, $N_p$-PAM modulation is briefly reviewed. In the sequel, the \acp{BEP} are derived considering a single carrier transmission of $N_p$-PAM symbols over \ac{AWGN} as well as {frequency-flat} Rayleigh channels.

\subsection{PAM Modulation}
By deploying \ac{PAM}, data symbols are constrained to the  alphabet
\begin{equation}
	\mathcal{A}_{N_p} = 
	\cbrac{-N_p +1, -N_p  +3,\cdots, N_p -1},
	\label{eq:pam_alphabet}
\end{equation}
meaning that only the in-phase component is deployed. Furthermore, one can easily infer that each $N_p$-PAM symbol can deliver
\begin{equation}
	N_b = \log_2(N_p)  
	\label{eq:pam_nb}
\end{equation}
bits by using an average symbol energy of 
\begin{equation}
	E_s = \dfrac{N_p^2-1}{3}.
	\label{eq:pam_energy}
\end{equation}

\subsection{{BEP for Single Carrier PAM Transmission over AWGN Channels}}
First, consider the simplest scenario: the transmission of PAM symbols over an \ac{AWGN} channel. Such a transmission can be modeled as 
\begin{equation}
	x = a + \eta,
	\label{eq:pam_model_awgn}
\end{equation}
where $a$ is the transmitted $N_p$-PAM symbol and $\eta$ is the additive noise described by the \ac{PDF}
\begin{equation}
f_{\eta}(y) = \dfrac{1}{\sqrt{\pi N_0}} \, e^{-\frac{y^2}{N_0}},
\end{equation}
where $N_0$ is the noise power density. Furthermore, one can rewrite the noise power in terms of the normalized \ac{SNR} $\gamma_b$\footnote{Bit energy per noise energy.} based on eq. \eqref{eq:pam_nb} and \eqref{eq:pam_energy}, \textit{i.e.},
\begin{equation}
	N_0 = \dfrac{N_p^2-1}{3\log_2 N_p \gamma_b}.
\end{equation}

A classical and straightforward approximation for evaluating the \ac{BEP} of \ac{PAM} symbols corrupted by additive noise is given by \cite{Proakis1996}:
\begin{equation}
P_b^\mt{awgn,pam}(\gamma_b) \approx \dfrac{2\rbrac{N_p-1}}{N_p\log_2N_p} 
                      \displaystyle\int_1^\infty f_\eta({z}) d{z}.
\label{eq:pam_ber_add_approx}
\end{equation}
Despite being very simple, eq. \eqref{eq:pam_ber_add_approx} only considers symbol errors caused by adjacent symbols. Moreover, eq. \eqref{eq:pam_ber_add_approx}  approximates the \ac{BEP} as the \ac{SEP} divided by the number of bits. {Thus, such expression considers that symbols are misestimated at the same rate of the bits.}

By solving eq. \eqref{eq:pam_ber_add_approx} considering an \ac{AWGN} channel, one can obtain the approximate {\ac{BEP}} of an $N_p$-\ac{PAM} transmission over an \ac{AWGN} channel:
\begin{equation}
	P_b^\mt{awgn,pam}(\gamma_b) \approx
	 \dfrac{2\rbrac{N_p-1}}{N_p\log_2N_p} 
	Q\rbrac{ \sqrt{\dfrac{6\log_2N_p}{N_p^2-1} \gamma_b } },
\label{eq:pam_ber_awgn_approx}
\end{equation}
where
\begin{equation}
Q(y)=
\dfrac {1}{\sqrt {2\pi }}
\displaystyle\int _{y}^{\infty }e^{-\frac {z^{2}}{2}}\,dz
\end{equation}
is the tail {distribution of a standard} normal distribution. Notice that an exact \ac{BEP} expression is far more complex as all error combinations must be taken into account. Fortunately, authors {of} \cite{Cho2002} offer the exact expression considering $N_p$-PAM Gray coded symbols:
\begin{align}
P_b^\mt{awgn,pam}(\gamma_b) \,\,= & \,\, \dfrac{1}{N_p	\log_2N_p} \times\\
     & \displaystyle\sum_{k=1}^{\log_2N_p} \,\,
      \displaystyle\sum_{i=0}^{(1-2^{-k})N_p-1}
      w_{i,k,N_p}
      \displaystyle\int_{2i+1}^\infty f_\eta({z}) d{z}, \notag
\label{eq:pam_ber_add_exact}
\end{align}
where
\begin{equation}
	w_{i,k,N_p}=
	(-1)^{ \floor{ i2^{k-1}/N_p} } 
	\rbrac{ 2^{k-1} - \floor{ \dfrac{i2^{k-1}}{N_p} + \dfrac{1}{2} } }.
\end{equation}
In particular, an \ac{AWGN} scenario leads to the exact \ac{BEP}
\begin{align}
P_b^\mt{awgn,pam}(\gamma_b) & =  \dfrac{2}{N_p	\log_2N_p} 
          \displaystyle\sum_{k=1}^{\log_2N_p}\,\,
          \displaystyle\sum_{i=0}^{(1-2^{-k})N_p-1}
          w_{i,k,N_p} \notag\\
&
\times \,\, Q\rbrac{ (2i+1) \sqrt{  \dfrac{6\log_2N_p}{N_p^2-1} \gamma_b }}.\label{eq:pam_ber_awgn_exact}
\end{align}

\subsection{{BEP for Single Carrier PAM Transmission over {Frequency-Flat} Rayleigh Channels}}
By considering that symbol transmission takes place over a Rayleigh channel, the transmission model presented in \eqref{eq:pam_model_awgn} must be rewritten as
\begin{equation}
	x = h a + \eta,
	\label{eq:pam_model_ray}
\end{equation}
where $h$ is the complex-valued channel coefficient.

Since the channel fades the signal randomly, one must consider all the possible values $h$ can take in order to evaluate the average \ac{BEP}. Thus, the overall \ac{BEP} is obtained by the integral
\begin{equation}
P_b^\mt{ray}(\gamma_b) = \displaystyle\int_0^{\infty} 
						 f_{\abs{h}^2}({z},\gamma_b) P_b^\mt{awgn}({z}) d{z},
\label{eq:pam_ber_ray}
\end{equation}
where
\begin{equation}
	f_{\abs{h}^2}({y},\gamma_b) = \dfrac{1}{\gamma_b} e^{-\frac{{y}}{\gamma_b}} u({y})
	\label{eq:pdf_h2}
\end{equation}
is the \ac{PDF} of the squared channel coefficient for a given normalized \ac{SNR} $\gamma_b$.

By combining and solving eqs. \eqref{eq:pam_ber_ray} and \eqref{eq:pam_ber_awgn_approx}, it can be concluded that an approximation for the \ac{BEP} of $N_p$-PAM symbols transmitted over {frequency-flat} Rayleigh channels is given by 
\begin{equation}
	\def\temp{ \frac{3\log_2N_p}{N_p^2-1} \gamma_b }
	\hspace*{-2mm}
	P_b^\mt{ray,pam}(\gamma_b)  \approx 
	\dfrac{{N_p-1}}{N_p\log_2N_p} 
	\rbrac{ 1-\dfrac{\sqrt{\temp}}{\sqrt{\temp+1}} }.
\label{eq:pam_ber_ray_approx}
\end{equation}
Similarly, an exact formula {can be obtained through} eqs. \eqref{eq:pam_ber_ray} and \eqref{eq:pam_ber_awgn_exact}:
\begin{equation}
\def\temp{ \frac{3(2i+1)^2\log_2N_p}{N_p^2-1} \gamma_b }
\arraycolsep = 1.4pt
\hspace*{-2.5mm}
\begin{array}{rcl}
P_b^\mt{ray,pam}(\gamma_b) & = & \dfrac{1}{N_p	\log_2N_p} 
          \displaystyle\sum_{k=1}^{\log_2N_p}\,\,
          \displaystyle\sum_{i=0}^{(1-2^{-k})N_p-1}
          w_{i,k,N_p}\\
         && \times \rbrac{ 1-\dfrac{\sqrt{\temp}}{\sqrt{\temp+1}} },
\end{array}
\label{eq:pam_ber_ray_exact}
\end{equation}
which was proposed by \cite{Lopes2007}.

{
At this point, it is noteworthy mentioning that the expressions presented in this section can be modified to characterize \ac{OFDM} which is the baseline multiplexing scheme deployed in most multicarrier comparisons.  First, notice that \ac{OFDM} deploys $N_q$-\ac{QAM} symbols, while the \ac{BEP} expressions presented previously are valid for $M_p$-\ac{PAM}. This issue can be solved by using $N_p=\sqrt{N_q}$ in the previous equations, as \ac{QAM} can be interpreted as the independent transmission of \ac{PAM} symbols over the in-phase and quadrature components.  Moreover, \ac{OFDM} systems deploy \ac{CP}, which reduces the overall \ac{SNR} of the signal and increases the error rate. In this sense, the \ac{OFDM} \ac{BEP} can be expressed by
\begin{equation}
\hspace*{-6mm}
	\text
	{
	\scalebox{.84}
	{
	$
	\begin{array}{rcl}
P_b^\mt{awgn,ofdm}(\gamma_b) & = & \dfrac{2}{\sqrt{N_q}	\log_2\sqrt{N_q}} 
          \displaystyle\sum_{k=1}^{\log_2\sqrt{N_q}}\,\,
          \displaystyle\sum_{i=0}^{(1-2^{-k})\sqrt{N_q}-1}
          w_{i,k,\sqrt{N_q}} \\
&   &
\times \,\, Q\rbrac{ (2i+1) \sqrt{  \dfrac{3\log_2{N_q}}{N_q-1} \dfrac{M}{M+N_{cp}}\gamma_b }}
\end{array}
$
}
}
\label{eq:pe_ofdm}
\end{equation}
for \ac{AWGN} and by
\begin{equation}
\hspace*{-5mm}
\text
{
\scalebox{.93}
{
$
\def\temp{ \frac{3(2i+1)^2\log_2{N_q}}{2\rbrac{N_q-1}} \dfrac{M}{M+N_{cp}}\gamma_b }
\arraycolsep = 1.4pt
\hspace*{-2.5mm}
\begin{array}{rcl}
P_b^\mt{ray,ofdm}(\gamma_b) & = & \dfrac{1}{\sqrt{N_q}	\log_2\sqrt{N_q}} 
          \displaystyle\sum_{k=1}^{\log_2\sqrt{N_q}}\,\,
          \displaystyle\sum_{i=0}^{(1-2^{-k})\sqrt{N_q}-1}
          w_{i,k,\sqrt{N_q}}\\
         && \times \rbrac{ 1-\dfrac{\sqrt{\temp}}{\sqrt{\temp+1}} }
\end{array}
$
}
}
\label{eq:pe_ofdm_ray}
\end{equation}
for flat-frequency Rayleigh channels
}

\section{$N_p$-PAM {\ac{BEP}}: FBMC Case}\label{sec:analytic}
Based on the foundation provided previously, this section builds up the closed-form \ac{BEP} expressions for  \ac{FBMC} systems over \ac{AWGN} and {frequency-flat} Rayleigh channels.

\subsection{{BEP for FBMC Transmission over AWGN Channels}}
For the \ac{FBMC} case over \ac{AWGN}, one must combine the multiplexed signal with the effect of the additive noise, \textit{i.e.}, 
\begin{equation}
	x[k] = s[k] + \eta[k],
	\label{eq:fbmc_model_awgn}
\end{equation}
where $\eta[k]\sim\mathcal{CN}(0,N_0)$. Hence, the reconstruction of the $n_0$th symbol of the $m_0$th subcarrier can be proceeded by combining eqs. \eqref{eq:fbmc_model_awgn} and \eqref{eq:fbmc_est1}, \textit{i.e.},
\begin{equation}
	\hspace*{-3mm}
	\arraycolsep = 1.4pt
	\begin{array}{rcl}
	{x_{m_0,n_0} }
	& = & 
	\real{ \left\langle x[k]\,| p_{m_0,n_0}[k] \right\rangle }\\
	&   & 
	a_{m_0,n_0} \\
	&   & 
	+ \displaystyle\sum_{\substack{n\neq n_0\\m\neq m_0}} a_{m,n} \epsilon_{m,n}
	+ \real{ \left\langle \eta[k]\, |\, p_{m_0,n_0}[k] \right\rangle }\\
	& = &
	a_{m_0,n_0} + \epsilon_0 + \eta_0,
	\end{array}
	\label{eq:fbmc_model_awgn2}
\end{equation}
{where $\epsilon_0$ is the overall interference over $a_{m_0,n_0}$.} At this point, one can observe that the easiest option to derive the \ac{BEP} of an \ac{FBMC} signal under \ac{AWGN} signal is {achieved} by 
\begin{itemize}
	\item[i.] obtaining the \ac{PDF} of the noise combined with the interference ($\epsilon_0 + \eta_0$);
	\item[ii.] deploying eq. \eqref{eq:pam_ber_add_approx} for an approximate \ac{BER} expression or \eqref{eq:pam_ber_awgn_exact} for the exact BER formula derivation.
\end{itemize}

\subsubsection{{PDF of the Noise plus Interference} }
{By recalling that the resulting \ac{PDF} of the sum of two random variables is the convolution of both \acp{PDF}, \textit{i.e.}, }
\begin{equation}
	f_{a+b}(y) = \displaystyle\int_{-\infty}^{\infty} f_a(z)f_b(y-z) dz{.}
	\label{eq:pdf_xpy}
\end{equation}
Thus, since the \ac{PDF} of {the} $n$th \ac{PAM} symbol of the $m$th subcarrier is modeled by
\begin{equation}
	f_{a_{m,n}}({y}) = 
	\dfrac{1}{N_p}
	\displaystyle\sum_{a_{m,n}\in\mathcal{A}_{N_p}} 
	\delta\rbrac{{y}-a_{m,n}}
	\label{eq:pdf_fbmc_pam}
\end{equation}
and $\epsilon_{m,n}$ is {deterministic}, one may conclude that 
\begin{equation}
	f_{a_{m,n}\epsilon_{m,n}}({y}) = 
	\dfrac{1}{N_p}
	\displaystyle\sum_{a_{m,n}\in\mathcal{A}_{N_p}} 
	\delta\rbrac{{y}-a_{m,n}\epsilon_{m,n}}.
	\label{eq:pdf_fbmc_pam_interf}
\end{equation}
{Notice that the summation notation adopted in eqs. \eqref{eq:pdf_fbmc_pam} and \eqref {eq:pdf_fbmc_pam_interf} provides a more compact presentation. In this notation, the summation includes all the elements of the $N_p$-\ac{PAM} set described in eq. \eqref{eq:pam_alphabet}.}

According to eq. \eqref{eq:fbmc_model_awgn2}, the interference experienced at the receiver side ($\epsilon_0$) is the combination of adjacent symbols $a_{m,n}$ scaled by their respective interference elements $\epsilon_{m,n}$. Hence, a series of convolutions must be evaluated in order to evaluate the \ac{PDF} of $\epsilon_0$, which leads to
\begin{align}
	f_{\epsilon_0}({y}) = &
	\dfrac{1}{N_p^{\abs{\mathcal{E}}}}
	\displaystyle\sum_{a_{0,-2K+1}\in\mathcal{A}_{N_p}} 
	\cdots
	\displaystyle\sum_{a_{M-1,2K-1}\in\mathcal{A}_{N_p}} \notag\\
	& \delta\rbrac{{y}-\displaystyle\sum_{(m,n)\in\mathcal{E}} a_{m,n} \epsilon_{m,n}},
	\label{eq:pdf_fbmc_total_interf}
\end{align} 
{where the last summation includes all the elements of $\mathcal{E}$, which are defined in eq. \eqref{eq:interf_set}. Once more, such a notation was adopted in favor of a more compact presentation.} Notice that eq. \eqref{eq:pdf_fbmc_total_interf} has a large number of coefficients as it takes into account all possible combination of interfering elements $a_{m,n}\epsilon_{m,n}$. In this sense, we limited the last summation to the set $\mathcal{E}$ in order to ignore null interfering elements and decrease the complexity of the equation.

\begin{figure*}[!htbp]
\begin{equation}
	\text
	{
	\scalebox{.76}
	{
	$
	f_{\epsilon_0+\eta_0}({y}) = 
	\dfrac{1}{N_p^{\abs{\mathcal{E}}}}
	\sqrt{\dfrac{1}{2\pi}\dfrac{6\log_2N_p }{N_p^2-1} \gamma_b}
	\displaystyle\sum_{a_{0,-2K+1}\in\mathcal{A}_{N_p}} 
	\cdots
	\displaystyle\sum_{a_{M-1,2K-1}\in\mathcal{A}_{N_p}} 
	\exp
	\sbrac{
	-\dfrac{6\log_2N_p }{N_p^2-1} \gamma_b
	\rbrac{\dfrac{{y}-\displaystyle\sum_{(m,n)\in\mathcal{E}} a_{m,n} \epsilon_{m,n}}{2}}^2
	}
	$
	}
	}
	\label{eq:pdf_fbmc_noise_interf}
\end{equation}
\vspace*{4mm}
\begin{equation}
	\text
	{
	\scalebox{.76}
	{
	$
	P_b^\mt{awgn,fbmc}
	{(\gamma_b)}
	{\approx}
	 \dfrac{2\rbrac{N_p-1}}{N_p^{\abs{\mathcal{E}}+1}\log_2N_p} 
	\displaystyle\sum_{a_{0,-2K+1}\in\mathcal{A}_{N_p}} 
	\cdots
	\displaystyle\sum_{a_{M-1,2K-1}\in\mathcal{A}_{N_p}} 
	Q
	\sbrac
	{	
	\sqrt{\dfrac{6\log_2N_p }{N_p^2-1} \gamma_b}
	\rbrac{1-\displaystyle\sum_{(m,n)\in\mathcal{E}} a_{m,n} \epsilon_{m,n} }
	}
	$
	}
	}
	\label{eq:fbmc_ber_awgn_approx}
\end{equation}
\vspace*{4mm}
\begin{equation}
	\text
	{
	\scalebox{.76}
	{
	$
	P_b^\mt{awgn,fbmc}
	{(\gamma_b)}
	= \dfrac{2}{N_p^{\abs{\mathcal{E}}+1}\log_2N_p} 
	\displaystyle\sum_{a_{0,-2K+1}\in\mathcal{A}_{N_p}} 
	\cdots
	\displaystyle\sum_{a_{M-1,2K-1}\in\mathcal{A}_{N_p}} 
    \displaystyle\sum_{k=1}^{\log_2N_p}
    \displaystyle\sum_{i=0}^{(1-2^{-k})N_p-1}
    w_{i,k,N_p}
	Q
	\sbrac
	{	
	\sqrt{\dfrac{6\log_2N_p }{N_p^2-1} \gamma_b}
	\rbrac{2i+1-\displaystyle\sum_{(m,n)\in\mathcal{E}} a_{m,n} \epsilon_{m,n} }
	}
	$
	}
	}
	\label{eq:fbmc_ber_awgn_exact}
\end{equation}
\vspace*{4mm}
\begin{equation}
\def\mya{\dfrac{6\log_2N_p }{N_p^2-1} \gamma_b}
\def\myc{2i+1-\displaystyle\sum_{(m,n)\in\mathcal{E}} a_{m,n} \epsilon_{m,n}}
	\hspace*{-4.5mm}
	\text
	{
	\scalebox{.76}
	{
	$
	\hspace{-15mm} P_b^\mt{ray,fbmc}
	{(\gamma_b)}
	 \approx
	\dfrac{N_p-1}{N_p^{\abs{\mathcal{E}}+1}\log_2N_p} 
	\displaystyle\sum_{a_{0,-2K+1}\in\mathcal{A}_{N_p}} 
	\cdots
	\displaystyle\sum_{a_{M-1,2K-1}\in\mathcal{A}_{N_p}} 
    \sbrac
    {
    1-
    \rbrac{\myc}
    \sqrt
    {
    \dfrac
    {\mya}
    {\rbrac{\myc}^2\mya+1}
    }
    }
	$
	}
	}
	\label{eq:fbmc_ber_ray_approx}
\end{equation}
\vspace*{4mm}
\begin{equation}
\def\myc{2i+1-\displaystyle\sum_{(m,n)\in\mathcal{E}} a_{m,n} \epsilon_{m,n}}
\def\mya{\dfrac{6\log_2N_p }{N_p^2-1} \gamma_b}
	\text
	{
	\scalebox{.76}
	{
	$
	\begin{array}{rcl}
	\hspace{-10mm} P_b^\mt{ray,fbmc} 
	{(\gamma_b)}
	& =  & \dfrac{1}{N_p^{\abs{\mathcal{E}}+1}\log_2N_p} 
	\displaystyle\sum_{a_{0,-2K+1}\in\mathcal{A}_{N_p}} 
	\cdots
	\displaystyle\sum_{a_{M-1,2K-1}\in\mathcal{A}_{N_p}} 
    \displaystyle\sum_{k=1}^{\log_2N_p}
    \displaystyle\sum_{i=0}^{(1-2^{-k})N_p-1}
    	w_{i,k,N_p} \\
    	&   &
    	\times
		\sbrac
    	{
    	1-
    	\rbrac{\myc}
    	\sqrt
    	{
    	\dfrac
    	{\mya}
    	{\rbrac{\myc}^2\mya+1}
    	}
    	}
    \end{array}
    $
    }
    }
    \vspace*{4mm}
	\label{eq:fbmc_ber_ray_exact}
	\end{equation}
\hrulefill
\end{figure*}

It is noteworthy mentioning that the increasing number of elements on the \ac{PDF} of the summation of random variables can be observed in other random processes. For example, this behavior can also be observed on the Bates and Irwin-Hall distributions \cite{Marengo2017,Bradley2002}, which consists of the summation of an arbitrary number of uniform random variables.

Since $f_{\epsilon_0}(x)$ is composed exclusively by Dirac delta functions, convolving it with any other function is trivial as the Dirac Function is the neutral element for convolution operation. Thus, the overall \ac{PDF} of the noise plus interference can be expressed by \eqref{eq:pdf_fbmc_noise_interf}. Therefore, by combining eq. \eqref{eq:pdf_fbmc_noise_interf} with eq. \eqref{eq:pam_ber_add_approx}, the approximate \ac{BEP} for \ac{FBMC} signals over \ac{AWGN} is obtained as \eqref{eq:fbmc_ber_awgn_approx}. On the other hand, the exact \ac{BEP} expression is obtained by combining eqs. \eqref{eq:pdf_fbmc_noise_interf} and \eqref{eq:pam_ber_awgn_exact}, resulting in eq. \eqref{eq:fbmc_ber_awgn_exact}.
%

\subsection{{BEP for FBMC Transmission over {Frequency-Flat} Rayleigh Channels}}

The received signal of an \ac{FBMC} system operating over {frequency-flat} Rayleigh channels can be expressed as 
\begin{equation}
	x[k] = \displaystyle\sum_{\ell=0}^{\infty} h[\ell] s[k-\ell] + \eta[k]{,}
	\label{eq:fbmc_model_ray}
\end{equation}
where $h[k]$ is the channel impulse response, which fades the transmitted signal and may introduce frequency {selectivity}. However, let us consider the simplifying assumption that each subchannel is frequency-flat. Such an assumption is plausible as multicarrier systems are typically designed to enable a simple one-tap per-subchannel equalization. {Notice, however, that} multi-tap equalizers \cite{Ihalainen2005,Waldhauser2008b,Ihalainen2011} can be deployed as an alternative in scenarios where the number of subcarriers is insufficient to lead to subchannel frequency-flatness. 

Aiming to retrieve the conveyed data, the received signal is pre-processed using the appropriate \ac{FBMC} pulse, leading to
\begin{equation}
	\begin{array}{rcl}
	x_{m_0,n_0} 
	& = & 
	\real{ \left\langle x[k]\, |\, p_{m_0,n_0}[k] \right\rangle }\vspace{2mm}\\
	& = & 
	H_{m_0} 
	\rbrac{
	a_{m_0,n_0} 
	+ \displaystyle\sum_{\substack{n\neq n_0\\m\neq m_0}} a_{m,n} \epsilon_{m,n}
	}	
	\vspace{2mm}\\
	&& + \,\, \real{ \left\langle \eta[k]\, |\, p_{m_0,n_0}[k] \right\rangle } \vspace{2mm}\\
		& {\approx} &
	H_{m_0}\rbrac{ a_{m_0,n_0} + \epsilon_0} + \eta_0,
	\end{array}
	\label{eq:fbmc_model_ray2}
\end{equation}
where $H_{m_0}$ is the frequency response of the $m_0$th subcarrier. {Notice that eq. \eqref{eq:fbmc_model_ray2} is an approximation, as one {can} recall that the subchannels were considered frequency-flat.} Moreover, eq. \eqref{eq:fbmc_model_ray2} is composed of the faded symbol plus the interference and the additive noise. From this perspective, the {approximate} \ac{BEP} expression for an \ac{FBMC} system operating over {frequency-flat} Rayleigh channels can be obtained by integrating eq. \eqref{eq:pam_ber_ray} considering eq. \eqref{eq:fbmc_ber_awgn_approx}, which leads to eq. \eqref{eq:fbmc_ber_ray_approx}, while the exact expression emerges from the integration of eq. \eqref{eq:pam_ber_ray} considering eq. \eqref{eq:fbmc_ber_awgn_exact}, yielding \eqref{eq:fbmc_ber_ray_exact}.

\section{Numerical Results}\label{sec:results}
This section presents the numerical results aiming to demonstrate the effectiveness of the \ac{BEP} expressions derived throughout this paper. In order to achieve such a goal, we compare the derived \ac{BEP} expressions with the simulated \ac{BER} for the \ac{OFDM} and \ac{FBMC} setups described in Table \ref{tab:fbmc_par}, considering both \ac{AWGN} and {frequency-flat} Rayleigh channels. {For \ac{FBMC}, two prototype {filters} were deployed, the \ac{EGF} and the Mirabbasi-Martin filter.} Notice that the \ac{EGF} \cite{Siohan2002} was chosen as the prototype filter due to its flexibility, which enables achieving different self-interference levels by tuning the spreading factor $\alpha$. This feature is useful to test the effectiveness of the derived expressions under very distinct scenarios. In particular, the \ac{EGF} set with $\alpha=1.00$ provides a large SIR of $60.49$ dB,  enabling near-perfect symbol reconstruction. On the other hand, by setting $\alpha=0.25$, the \ac{SIR} of the \ac{EGF} is reduced to $21.27$ dB, leading to a poor \ac{BER} performance. Moreover, the Mirabbasi-Martin prototype filter is also deployed in this comparison as it presents a high spectrum and symbol reconstruction performances. Indeed, such a prototype filter is one of the most popular choices, being recommended in the PHYDIAS project \cite{Phydias}.

\begin{table}[!htbp]
	\centering
	\caption{Parameters for the \ac{FBMC} system.}
	\begin{tabular}{rllll}
		\hline
		\textbf{Parameter} & & \textbf{OFDM} & & \textbf{FBMC}
		\\\hline
		Modulation order            & & $64$-QAM & & $8$-PAM
		\\
		Subcarriers, $M$            & & $16$     & & $16$ 
		\\
		Cyclic Prefix               & & $M/8$    & & $0$
		\\
		Overlapping factor, $K$     & & -        & & $4$ 
		\\
		Filter length, $L_p$        & & -        & & $MK+1$ 
		\\
		
		Prototype Filter            & & - & & EGF, $\alpha=\cbrac{0.25,1.00}$,
		\\
		                            & &   & & Mirabbasi-Martin
		\\	
		\hline
	\end{tabular}
	\label{tab:fbmc_par}
\end{table}

\subsection{Complexity Considerations}
\begin{figure}[!htbp]
	\centering
	\includegraphics[width=.65\textwidth]{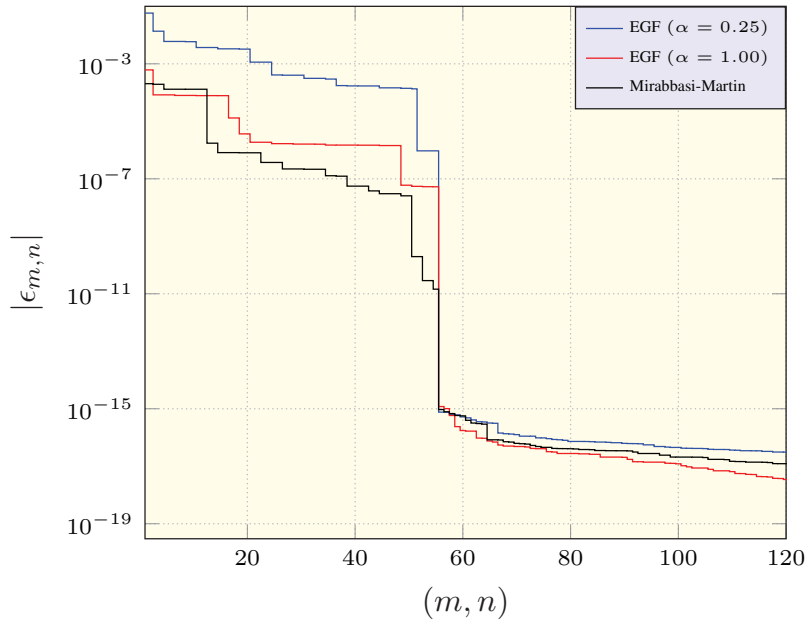}
	\caption{Ordered absolute values for the interference elements $\epsilon_{m,n}$ for the \ac{FBMC} system described in Table \ref{tab:fbmc_par}.}
	\label{fig:fbmc_e}
\end{figure}

Before presenting the \ac{BEP} results, let us discuss the complexity of the derived expressions. Despite not relying on complex functions, the main issue of the derived expressions is the large number of terms of the summations. A more careful analysis reveals that the {\ac{BEP}} expressions in eq. \eqref{eq:fbmc_ber_awgn_approx}, \eqref{eq:fbmc_ber_awgn_exact}, \eqref{eq:fbmc_ber_ray_approx} and \eqref{eq:fbmc_ber_ray_exact} presents at least $N_p^{\abs{\mathcal{E}}+1}$ terms. Hence, the summation of a huge number of terms is required in order to provide the theoretical \ac{BEP} expression. As an example, evaluating the \ac{BEP} of the reasonably small system portrayed in Table \ref{tab:fbmc_par} would require the evaluation of approximately $3 \cdot10^{107}$ terms. However, the interference elements  $\epsilon_{m,n}$ decay rapidly as shown in Fig. \ref{fig:fbmc_e}. Therefore, let us restrain the set $\mathcal{E}$ to the $8$ largest elements in order to make the developed expressions feasible {and therefore useful}. {Notice that even} with the limited set, the theoretical expressions require the summation of, at least, $16\cdot10^{6}$ elements. 

\subsection{\ac{BER} Performance}
\begin{figure}[!htbp]
\centering
\includegraphics[width=\textwidth]{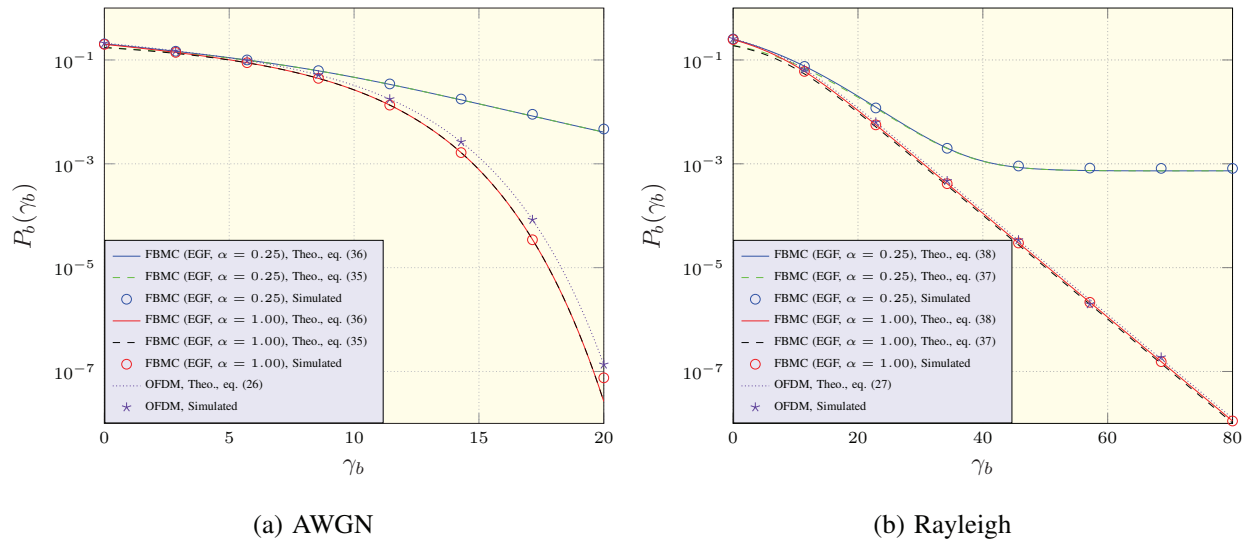}
\caption{Theoretical and simulated \ac{BER}  for the \ac{FBMC} system described in Table \ref{tab:fbmc_par} and with \ac{EGF} prototype filter.}
\label{fig:fbmc_ber_theo_awgn_ray_egf}
\end{figure}

{In Figures \ref{fig:fbmc_ber_theo_awgn_ray_egf}.(a) and \ref{fig:fbmc_ber_theo_awgn_ray_egf}.(b), we present both the simulated \ac{BER} and the theoretical \ac{BEP} for the \ac{FBMC} system described in Table \ref{tab:fbmc_par} considering an \ac{EGF} prototype filter, and \ac{AWGN} and {frequency-flat} Rayleigh scenarios}. With $\alpha=1.00$ the \ac{FBMC} system showed no apparent performance loss, whereas with $\alpha=0.25$, the \ac{BER} performance is considerably severed, leading to a \ac{BER} floor around $10^{-3}$ in  the Rayleigh scenario. Such behavior is established due to the degradation of the prototype filter \ac{SIR}, which dictates the amount of self-interference generated by the prototype filter. However, {it is} important to state that the prototype filter \ac{SIR} improves as $\alpha$ increases, while the spectrum gets increasingly poor. As an example, \acp{EGF} with $\alpha$ values of $0.5$, $1$ and $2$ lead to \ac{SIR} levels of $33.73$,  $60.49$ and $114.48$ dB, respectively, whereas the \ac{OoB} energy emission increases gradually to $-33.95$, $-19.69$ and $-12.46$ dB \cite{Kobayashi2019}.

\begin{figure}[!htbp]
	\centering
	\includegraphics[width=\textwidth]{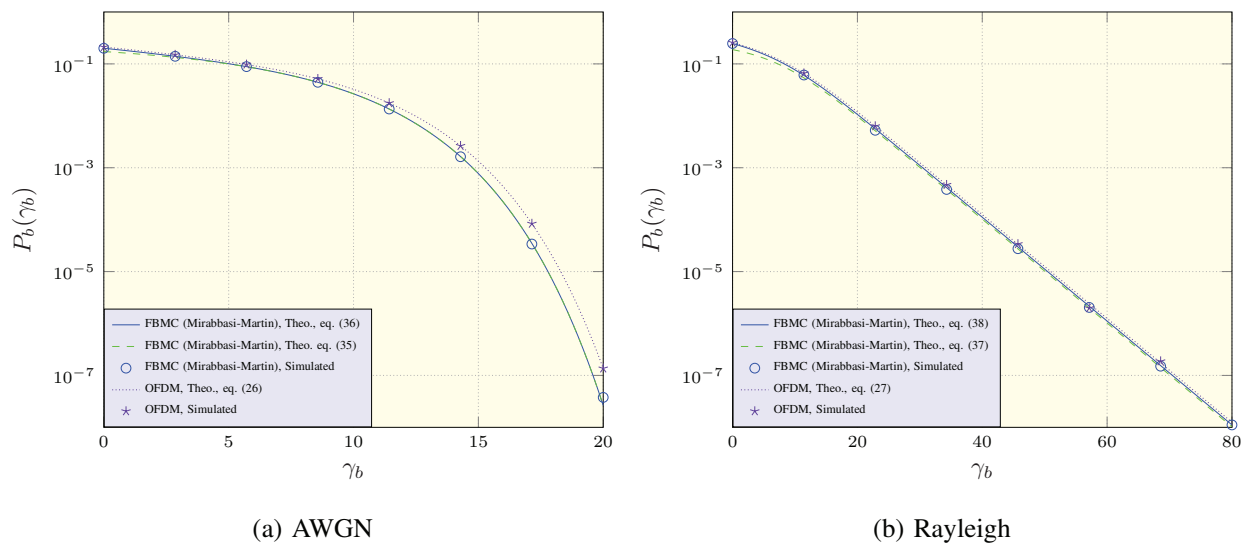}
	\caption{Theoretical and simulated \ac{BER} for the \ac{FBMC} system described in Table \ref{tab:fbmc_par} and with Mirabbasi-Martin prototype filter.}
\label{fig:fbmc_ber_theo_awgn_ray_martin}
\end{figure}

{
In order to prove the effectiveness of the derived expression for any prototype filter, Figures \ref{fig:fbmc_ber_theo_awgn_ray_martin}.(a) and \ref{fig:fbmc_ber_theo_awgn_ray_martin}.(b) {compare} the \ac{BER} with the \ac{BEP} for an \ac{FBMC} system deploying a Mirabbasi-Martin prototype filter \cite{Mirabbasi2003} and the parameters described in Table \ref{tab:fbmc_par}. Once more, the simulation  and the analytically  results depicted in Figures \ref{fig:fbmc_ber_theo_awgn_ray_martin}.(a) and \ref{fig:fbmc_ber_theo_awgn_ray_martin}.(b) are very close, proving the effectiveness of the proposed \ac{BEP} expressions. Moreover, since the Mirabbasi-Martin prototype filter presents a high \ac{SIR} level, {\it i.e.,} $\approx65.25$ dB, performance losses were not observed in Figures \ref{fig:fbmc_ber_theo_awgn_ray_martin}.(a) and \ref{fig:fbmc_ber_theo_awgn_ray_martin}.(b).
}

Regarding the effectiveness of the derived formulas, one can observe that such expressions are very close to the simulation results. Despite reducing the size of the set $\mathcal{E}$ considerably, such approximation led a small influence on the overall results, as $\epsilon_{m,n}$ elements decay rapidly. Thus, the provided expression demonstrated to be effective to predict the \ac{BER} performance of \ac{FBMC} systems operating over \ac{AWGN} and Rayleigh channels, regardless of the prototype filter choice.

Additionally, one can observe that the numerical results for \ac{OFDM} match with the {values found by the} \ac{BEP} expressions provided in section \ref{sec:Pe-SC}. Compared to \ac{FBMC} with \ac{EGF} with $\alpha=1$ and Mirabbasi-Martin prototype filters, the \ac{BER} performance of \ac{OFDM} presents a small offset due to the {\ac{CP} deployment,} which does not carry actual data, reducing the overall \ac{SNR} of the signal. On the other hand, prototype filters with low symbol reconstruction performance, \textit{e.g.}, \ac{EGF} with $\alpha=0.25$, can limit the \ac{BER} performance of \ac{FBMC} systems operating under high \ac{SNR} regime as observed in Figure \ref{fig:fbmc_ber_theo_awgn_ray_egf}.(b).

\section{Conclusions}\label{sec:concl}
Throughout this work, we derived the theoretical \ac{BEP} expressions for \ac{FBMC} systems under \ac{AWGN} and {frequency-flat} Rayleigh channels, regardless of the prototype filter choice. Despite the required computational burden processing, the provided formulas are exact. Fortunately, the derived expressions  have resulted {relatively} accurate yet by restricting the interference set $\mathcal{E}$ to the most significant values. {Thus, such an approximation} does not compromise the final BEP prediction values. From this perspective, the numerical results demonstrate that the proposed \ac{BEP} expressions can {adequately} characterize the \ac{BER} performance of \ac{FBMC} systems. Thus, this work quantifies the effect of the prototype filter on the overall \ac{BER} performance of \ac{FBMC} systems, which until now was only known qualitatively in terms of dependencies on \ac{SIR} and distortion measurements.

\section*{Acknowledgments}
This study was financed in part by the National Council for Scientific and Technological Development (CNPq) of Brazil under grants 404079/2016-4 and 304066/2015-0, {in part by the CAPES - Brazil - Finance Code 001 (PhD scholarship)}, and in part by Londrina State University, Parana State Government.

\cleardoublepage

\section*{List of Acronyms}
\begin{itemize}[ leftmargin = 5em,
			     labelsep   = .75em,
                 font       = \bfseries
                 ]
\setlength\itemsep{.3em}
\small
\item[BER] Bit Error Probability
\item[CP] Cyclic Prefix
\item[DFE] Decision Feedback Equalizer
\item[DPSS] Discrete Prolate Spheroidal Sequences
\item[DTFT] Discrete-Time Fourier Transform
\item[EGF] Extended Gaussian Function
\item[FBMC] Filter Bank Multi-Carrier
\item[GFDM] Generalized Filtered-Division Filtered Multi-Carrier
\item[IB] In-Band
\item[ICI] Inter-Carrier Interference
\item[ISI] Inter-Symbol Interference
\item[MMSE] Minimum Mean-Square-Error
\item[MSA] Minimum Stopband Attenuation
\item[OFDM] Orthogonal Frequency-Division Multiplexing
\item[OFDP] Optimal Finite Duration Pulses
\item[OOB] Out-of-Band
\item[PAM] Pulse Amplitude Modulation
\item[PAPR] Peak-to-Avarage Power Ratio
\item[QAM] Quadrature Amplitude Modulation
\item[QCQP] Quadratically Constrained Quadratic Program 
\item[QP] Quadratic Program
\item[RMS] Root Mean Square
\item[SDP] SemiDefinite Program
\item[SIR] Signal-to-Interference Ratio
\item[SNR] Signal-to-Noise Interference
\item[SOCP] Second Order Conic Program 
\item[SRRC] Square-Root Raised-Cosine
\item[UFMC] Universal Filtered Multi-Carrier
\item[V-BLAST] Vertical Bell Laboratories Layered Space-Time
\end{itemize}

\begin{acronym}[\hspace*{3.3cm}] 
\acro{4G}{4th Telecommunication Generation}
\acro{5G}{5th Telecommunication Generation}
\acro{AWGN}{Additive White Gaussian Noise}
\acro{BER}{Bit Error Rate}
\acro{BEP}{Bit Error Probability}
\acro{SEP}{Symbol Error Probability}
\acro{BS}{Base Station}
\acro{CP}{Cyclic-Prefix}
\acro{cMBB}{crowd Mobile BroadBand}
\acro{DPSS}{Discrete Prolate Spheroidal Sequences}
\acro{DTFT}{Discrete-Time Fourier Transform}
\acro{EGF}{Extended Gaussian Function}
\acro{eMBB}{enhanced Mobile BroadBand}
\acro{FBMC-OQAM}{Offset Quadrature Amplitude Modulation Filter Bank MultiCarrier}
\acro{FBMC}{Filter Bank MultiCarrier}
\acro{FFT}{Fast Fourier Transform}
\acro{GFDM}{Generalized Frequency-Division Multiplexing}
\acro{IAM}{Interference Approximation Method}
\acro{IB}{In-Band}
\acro{ICI}{InterCarrier Interference}
\acro{IOTA}{Isotropic Orthogonal Transform Algorithm}
\acro{ISI}{InterSymbol Interference}
\acro{ITU}{International Telecommunication Union}
\acro{IoT}{Internet of Things}
\acro{KKT}{Karush-Kuhn-Tucker}
\acro{LTE}{Long-Term Evolution}
\acro{M2M}{Machine-to-Machine}
\acro{MIMO-FBMC}{Multiple-Input Multiple-Output Filter Bank MultiCarrier}
\acro{MIMO}{Multiple-Input Multiple-Output}
\acro{MMSE-DFE}{Minimum Mean-Square Error Decision Feedback Equalizer}
\acro{MMSE}{Minimum Mean-Square Error}
\acro{mMTC}{massive Machine-Type communications}
\acro{mmWaves}{Millimeter Waves}
\acro{MSL}{Maximum Sidelobe Level}
\acro{MTC}{Machine-Type communications}
\acro{OFDM-OQAM}{Offset Quadrature Amplitude Modulation Frequency-Division Multiplexing}
\acro{OFDM}{Orthogonal Frequency-Division Multiplexing}
\acro{OFDP}{Optimal Finite Duration Pulses}
\acro{OQAM}{Offset Quadrature Amplitude Modulation}
\acro{OoB}{Out-of-Band}
\acro{PAM}{Pulse Amplitude Modulation}
\acro{PAPR}{Peak-to-Average Power Ratio}
\acro{POP}{Pair of Pilots}
\acro{PSK}{Phase-shift keying}
\acro{QAM}{Quadrature Amplitude Modulation}
\acro{QCQP}{Quadratically Constrained Quadratic Programming}
\acro{RAP}{Random Access Protocol}
\acro{RC}{Raised-Cosine}
\acro{RMS}{Root Mean Square}
\acro{SAR}{Specific Absorption Rate}
\acro{SDP}{SemiDefinite Programming}
\acro{SIR}{Signal-to-Interference Ratio}
\acro{SISO}{Single-Input Single-Output}
\acro{SNR}{Signal-to-Noise Ratio}
\acro{SOCP}{Second-Order Cone Program}
\acro{SRRC}{Square-Root Raised Cosine}
\acro{SUCR}{Strongest-User Collision Resolution}
\acro{UFMC}{Universal Filtered MultiCarrier}
\acro{URLLC}{Ultra-Reliable Low Latency Communications}
\acro{V-BLAST}{Vertical Bell Labs Layered Space-Time}
\acro{Wimax}{Worldwide Interoperability for Microwave Access}
\acro{ZF}{Zero-Forcing}
\acro{PDF}{Probability Density Function}
\acro{CDF}{Cumulative Density Function}

\end{acronym}

\end{document}